\documentclass[a4paper,useAMS,usenatbib]{mnras}
\usepackage{graphics,epsfig,psfig}
\usepackage{todonotes}
\usepackage[normalem]{ulem}
\usepackage{xcolor}
\usepackage{float}
\usepackage[style=base, singlelinecheck=off, font=small, labelfont=bf, 
justification=raggedright]{caption}
\usepackage{subfigure}
\usepackage{graphicx}
\usepackage[]{inputenc,amssymb}
\usepackage{natbib}
\usepackage{url}
\usepackage{widetext}
\usepackage{amsmath}
\usepackage{cancel}

  \makeatletter
\def\@fnsymbol#1{\ensuremath{\ifcase#1\or \dagger\or \ddagger\or
   \mathsection\or \mathparagraph\or \|\or **\or \dagger\dagger
   \or \ddagger\ddagger \else\@ctrerr\fi}}
    \makeatother
\def \be{\begin{equation}}
\def \ee{\end{equation}}
\def \bea{\begin{eqnarray}}
\def \eea{\end{eqnarray}}

\def\apj{ApJ}
\def\mnras{MNRAS}

\def\aap{A\&A}
\definecolor{webgreen}{rgb}{0,.5,0}
\definecolor{webbrown}{rgb}{.6,0,0}

\def\aap{A\&A}
\def\apj{ApJ}
\def\apjs{ApJS}
\def\apjl{ApJL}
\def\mnras{MNRAS}

\def\aj{AJ}

\def\pasp{PASP}              

\def\jcap{JCAP}
\def\sovast{Soviet Astronomy}

\title[Can we distinguish ABH from PBH?]{Can we distinguish astrophysical from primordial black holes via the stochastic gravitational wave background?}
\author[Mukherjee \& Silk (2021)]{Suvodip Mukherjee$^{1,2,3}$\thanks{s.mukherjee@uva.nl},  Joseph Silk$^{4, 5, 6} $\thanks{silk@iap.fr}\\
$^{1}$ Gravitation Astroparticle Physics Amsterdam (GRAPPA),
Anton Pannekoek Institute for Astronomy and Institute for Physics,\\
University of Amsterdam, Science Park 904, 1090 GL Amsterdam, The Netherlands\\
$^2$ Institute Lorentz, Leiden University, PO Box 9506, Leiden 2300 RA, The Netherlands\\
$^3$Delta Institute for Theoretical Physics, Science Park 904, 1090 GL Amsterdam, The Netherlands\\
$^{4}$ Institut d'Astrophysique de Paris, UMR 7095, CNRS, Sorbonne Universit\'e, 98bis Boulevard Arago, 75014 Paris, France\\
$^{5}$ The Johns Hopkins University, Department of Physics \& Astronomy, 3400 N. Charles Street, Baltimore, MD 21218, USA\\
$^{6}$ Beecroft Institute for Cosmology and Particle Astrophysics, University of Oxford, Keble Road, Oxford OX1 3RH, UK\\
}
\pubyear{2021}
\begin{document}
\label{firstpage}
\pagerange{\pageref{firstpage}--\pageref{lastpage}}
\maketitle

\label{firstpage}

\begin{abstract}
One of the crucial windows for distinguishing astrophysical black holes from primordial black holes is through the redshift evolution of their respective merger rates. The low redshift population of black holes of astrophysical origin is expected to follow the star formation rate. The corresponding peak in their merger rate peaks at a redshift smaller than that of the star formation rate peak   ($z_p \approx 2$), depending on the time delay between the formation and mergers of black holes. Black holes of primordial origin are going to be present before the formation of the stars, and the merger rate of these sources at high redshift is going to be large. We propose a joint estimation of a hybrid merger rate from the stochastic gravitational wave background, which can use the cosmic history of merger rates to distinguish between the two populations of black holes. Using the latest bounds on the amplitude of the stochastic gravitational wave background amplitude from the third observation run of LIGO/Virgo, we obtain weak constraints at $68\%$ C.L. on the primordial black hole merger rate index   $2.56_{-1.76}^{+1.64}$ and astrophysical black hole time delay  $6.7_{-4.74}^{+4.22}$ Gyr. 
We should be able to distinguish between the different populations of black holes with the forthcoming  O5 and A+ detector sensitivities.
\end{abstract}


\section{Introduction}
How are black holes forming in the Universe? There are two channels to form black holes, namely the astrophysical channel from the deaths of stars and the formation of black holes in the early Universe from primordial fluctuations \citep{1967SvA....10..602Z,1971MNRAS.152...75H,1975ApJ...201....1C, 1980PhLB...97..383K, 1985MNRAS.215..575K, Carr:2005zd, Sasaki:2018dmp}. The discovery of primordial black holes (PBHs) would change our fundamental understanding of the Universe and address one of the long-standing questions in the standard model cosmology about the nature of dark matter \citep{Bird:2016dcv, Carr:2016drx}. Even though PBHs are not candidates for the bulk of the dark matter, at least in the solar mass range, there remains an interesting question about whether a population of PBHs exists in the Universe that contributes to the observed gravitational wave events. One of the primary challenges is to decide whether one may be able to distinguish between astrophysical black holes (ABHs) forming from stellar deaths and  PBHs arising from an early universe origin.

The discovery of gravitational waves \citep{Abbott:2016blz, TheLIGOScientific:2016htt} has opened a new window for detection of  PBHs by exploring the properties of the GW merger rates and associated source populations, such as the masses of individual sources and the spins of the GW sources \citep{Bird:2016dcv, Clesse:2016vqa, Sasaki:2016jop, Gow:2019pok, Jedamzik:2020ypm, Jedamzik:2020omx, DeLuca:2020jug}.  
Several recent studies have considered the properties of the mass distributions of GW sources from individual events in order to distinguish between  ABHs and PBHs \citep{Hall:2020daa, Wong:2020yig,2021JCAP...05..003D, 2021JCAP...03..068H, Franciolini:2021tla}. 
However such studies are vulnerable to freedom in higher generation merging models that enable the upper mass gap, a powerful indicator of first-generation ABHs,  to be bridged \citep{Rodriguez:2019huv, 2020arXiv201105332K, Hamers:2021olp}.

Perhaps the ultimate signature that can potentially distinguish between different formation channels of black holes is through the evolution of the merger rate with redshift \citep{Raidal:2017mfl, Raidal:2018bbj, Vaskonen:2019jpv, Atal:2020igj, DeLuca:2020qqa}. GW sources produced from the deaths of stars form after the formation of the first stars, whereas PBHs exist in large numbers at a very high redshift before any stars formed. This is one of the key differences that can be used to distinguish between a population of ABHs and PBHs. For the ABHs, even though one would expect the merger rate of black holes to be related to the star formation rate of the Universe, one of the major sources of uncertainty in the merger rate is due to the time delay between the formation and merger of GW sources \citep{2012ApJ...759...52D, Dominik:2014yma, Lamberts:2016txh, Dvorkin:2016wac, Eldridge:2018nop, Vitale:2018yhm, Safarzadeh:2020qru, Santoliquido:2020axb}. For  GW sources with small-time delays ($t^{eff}_d < 100$ Myr), the peak of the mergers of  GW sources is around the peak of the star formation rate ($z\approx 2$), whereas for the scenarios with larger time delays, the peak of the merger of the ABHs can be shifted towards lower redshifts. However for PBHs, {\it the merger rate is always an increasing function of redshift}.  The number of mergers at high redshift for GW sources of primordial origin is always going to surpass the ABH merger rate \citep{Ali-Haimoud:2017rtz, Raidal:2017mfl, Raidal:2018bbj, Vaskonen:2019jpv, Atal:2020igj, DeLuca:2020qqa}. The formation of PBHs can also give additional sources of stochastic GW background \citep{Kohri:2018awv, Espinosa:2018eve, Wang:2019kaf}.  

This distinction may be largely academic since GW sources at high redshifts cannot be detected as  individual events (apart from rare  lensed events \citep{PhysRevLett.80.1138, Wang:1996as,Dai:2016igl, Broadhurst:2018saj, Broadhurst:2019ijv, Oguri:2019fix,Mukherjee:2020tvr}). However the high merger rate of the GW sources at high redshift leads to a stochastic gravitational wave background due to the contribution from  unresolved sources \citep{Allen:1996vm,Phinney:2001di,Regimbau:2007ed,Wu:2011ac,Romano:2016dpx,TheLIGOScientific:2016wyq,Abbott:2017xzg,LIGOScientific:2019vic,Abbott:2021xxi}.  We show that {\it PBH  mergers contribute to a potentially detectable stochastic GW background signal} \citep{Wang:2016ana, Mandic:2016lcn}. 

Measurement of the stochastic GW background (even in the absence of a detection) can probe the high redshift merger rate and its evolution with redshift \citep{10.1093/mnras/stz3226, Boco:2019teq, Callister:2020arv}. Using data from the third observational run, LIGO/Virgo has estimated the stochastic GW background power spectrum \citep{Abbott:2021xxi}. Though the measurement has not detected the stochastic GW background, it has provided an upper bound on the stochastic GW background power spectrum\citep{Abbott:2021xxi}.  

Here we construct a hybrid merger rate model of ABHs and PBHs,   by taking into account the time delay between formation and mergers of the astrophysical sources and incorporating a general redshift dependence of the PBH merger rate. Our hybrid model 
is driven by the use of the Madau-Dickinson star formation rate history (SFR) to derive the ABH merger rate and a power-law model for the PBH merger rate. The free parameters of this model are the time delay parameter, the local merger rate, the index of the power-law model of PBH merger rate, the characteristic mass-scale of PBHs, and the fraction of PBHs over ABHs.  This five-parameter model makes it possible to perform a relatively fast MCMC search of GW sources to jointly probe the parameter space of ABHs and PBHs. In this paper, we show the current constraints on the parameters of this hybrid model of the merger rate from the LIGO/Virgo third observing run of the stochastic GW background \citep{Abbott:2021xxi} and also using the bounds on the local merger rate from the individual events of the first half of the third observation run of O3a \citep{Abbott:2020niy, Abbott:2020gyp}. We also provide forecasts for the O5 observation run of the LIGO/Virgo detectors in its design sensitivity \citep{TheLIGOScientific:2014jea, TheVirgo:2014hva} and for the enhanced A+ sensitivity\citep{Aasi:2013wya, aplus}.
We do not consider the possible contributions from population-II/population-III astrophysical sources at high redshift, as there is no detection of the stochastic GW background from O3 observations. The method proposed here can easily be extended to include the contribution from such sources \citep{Inayoshi:2021atf}.

\section{Hybrid model of the merger rate for ABHS and PBHs}
We consider a parametric hybrid model of GW merger rates as a function of redshift to search for ABHs and PBHs. The model is composed of these two components with  corresponding probability distributions of  masses $P_{ABH}(m_i)$ and $P_{PBH}(m_i)$  written as 
\begin{align}\label{rateabhpbh}
\begin{split}
    \mathcal{R}_{GW}(z_m, m_1, m_2)= & \mathcal{N}\bigg[P_{ABH}(m_1)P_{ABH}(m_2)\mathcal{R}_{ABH}(z_m,m_2, m_2)\\ &  + P_{PBH}(m_1)P_{PBH}(m_2)\mathcal{R}_{PBH}(z, m_1, m_2)\bigg],
    \end{split}
\end{align}
where $\mathcal N$ is the normalization factor such that the local merger rate integrated over the mass distribution agrees with the value inferred from the individual detected events. $\mathcal{R}_{ABH}(z_m,M)$ is the merger rate for the ABHs that are expected to follow the cosmic star formation rate history. For the low redshift universe, we  assume that  the Madau-Dickinson fitting form \citep{Madau:2014bja} is a proxy for the star formation rate history. We  write the model of the  ABHs as 
\begin{align}\label{rate1}
\begin{split}
    \mathcal{R}_{ABH}(z_m,m_1, m_2)= \int^{\infty}_{z_m} dz& \frac{dt_f}{dz}R_{ABH}(0,m_1,m_2)\\& \, \times P(t^{eff}_d, m_1, m_2)  R_{SFR}(z),
    \end{split}
\end{align}
where $R_{SFR}(z)$ is the star-formation rate  motivated by the Madau-Dickinson relation \citep{Madau:2014bja}
\begin{equation}\label{mdsfr}
R_{SFR}(z)\propto \frac{(1+z)^{2.7}}{1+ (\frac{(1+z))}{2.9})^{5.6}}.
\end{equation}
In Eq. \ref{rate1}, the time-delay distribution between the formation and mergers of the ABHs is denoted by $P(t^{eff}_d, M)$, where $t^{eff}_d$ denotes the time-delay between formation and mergers. The redshift evolution of the ABH merger rate denoted in Eq. \eqref{rate1} is solely decided by the time-delay model. Depending on the probability distribution of the time delay, the peak of the merger rate, the slope of the merger rate at low redshift, and the slope of the merger rate at high redshift are specified. The only degree of freedom in this model is considered to be the adopted time-delay distribution. Currently, we have very little idea of the value of the time-delay parameter. Stellar population synthesis models suggest that the mergers of the different kinds of GW sources can be delayed by as much as a few hundreds of Myr up to about the age of the Universe \citep{2012ApJ...759...52D, Dominik:2014yma, Lamberts:2016txh, Eldridge:2018nop, Santoliquido:2020axb}. 

\begin{figure}
    \centering
    \includegraphics[width=1.\linewidth]{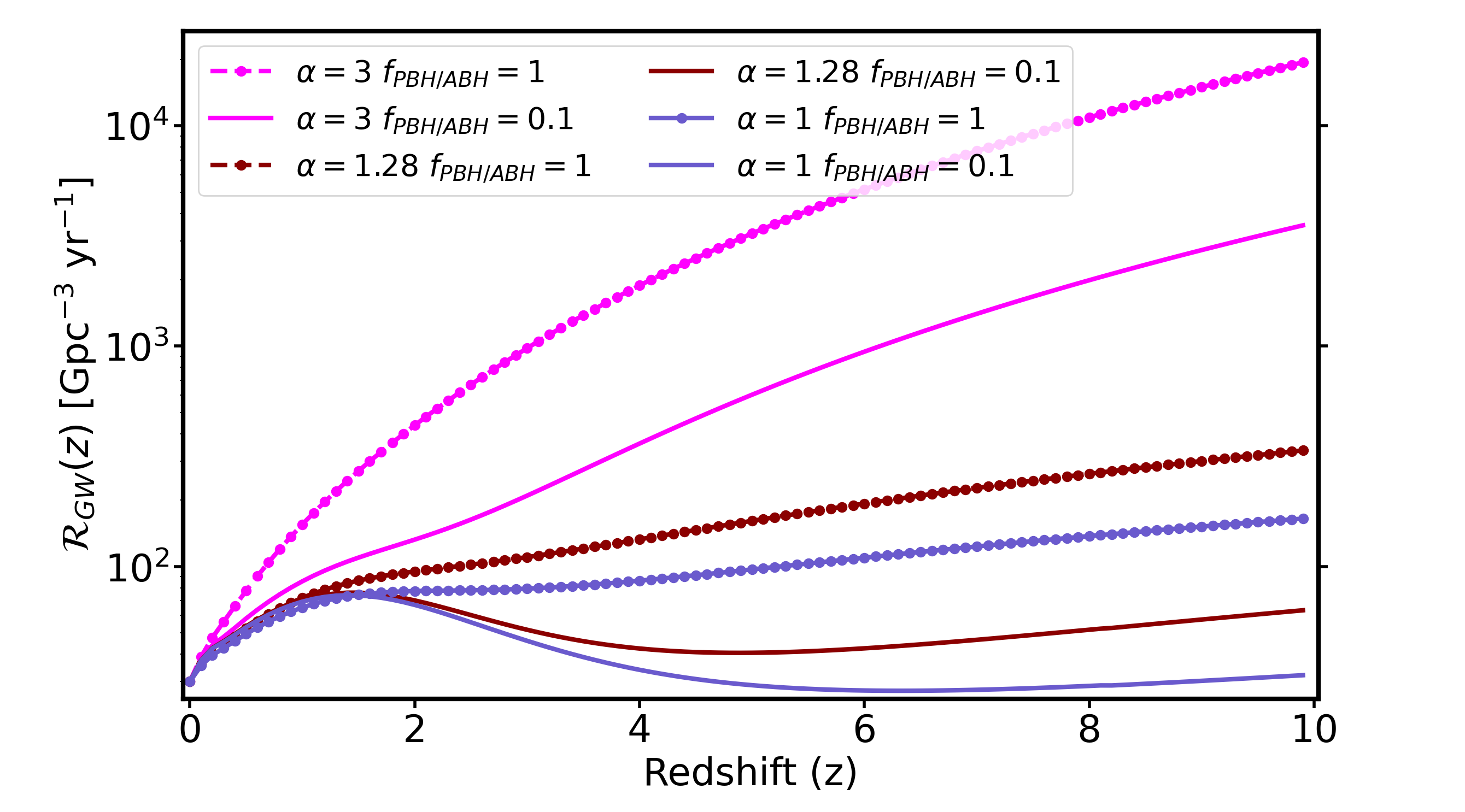}
    \caption{Hybrid model merger rates with a  minimum time-delay value $t^{eff}_d= 100$ Myr and  for different power-law indices $\alpha$ of the redshift dependence of the  PBH populations and  values of the fractional PBH abundance with respect to the ABH abundance $f_{PBH/ABH}$.}
    \label{fig:mergerrate}
\end{figure}

The second term in Eq. \eqref{rateabhpbh} captures the redshift evolution of the PBHs. The merger rate of PBHs can be modeled depending on whether the binaries are dominated by Poisson statistics or whether there is clustering. In the presence of strong clustering, the current merger rate is exponentially decreased, even for a large fraction of PBHs as dark matter. The redshift evolution of the merger rate is one of the key signatures for distinguishing between the clustered and Poissonian scenarios. The general behavior of the PBH merger rate is that there is an increase in the merger rate with increasing redshift, and so we model the PBH merger rate as
\begin{equation}\label{pbhsfr}
R_{PBH}(z)= R_{PBH}(0, m_1,m_2)(1+z)^\alpha,
\end{equation}
where $\alpha$ is a positive index in the power-law model and $R_{PBH}(0, m_1, m_2)$ denotes the local merger rate of the GW sources of primordial origin.  {We consider $\alpha$ as a free parameter to search for PBHs in a model-independent way from the stochastic GW background. However, for most of the known scenarios of black formation, the value of $\alpha \sim 1.3$ for the Poisson distribution \citep{ Raidal:2017mfl, Sasaki:2018dmp, Raidal:2018bbj}. For the forecast studies in the latter part of the paper, we consider the fiducial value of $\alpha=1.3$.} The merger rate of the GW sources is degenerate between the clustering signal and the PBH fraction  \citep{Raidal:2017mfl, Raidal:2018bbj, Young:2019gfc, Vaskonen:2019jpv, Atal:2020igj, DeLuca:2020qqa}. If the spatial clustering $\xi_{PBH}>> 1$ is very large, then the local merger rate can be exponentially suppressed, even if the fraction of PBH in dark matter $f_{PBH}=1$. We can express the local merger rate for the extremely clustered scenario as 
\begin{align}\label{pbhrate}
\begin{split}
R_{PBH}(z=0) \propto\, & \xi^{0.7}_{PBH}f^{1.7}_{PBH}\exp{(-(\xi_{PBH}f_{PBH}/10^4))}, \\ & \text{for\,\,} \xi_{PBH}f_{PBH}>10^3.
\end{split}
\end{align}
As a result, the relatively 
low observed merger rate of GW sources does not necessarily imply that $f_{PBH} < 10^{-2}$
\citep{Raidal:2017mfl, Vaskonen:2019jpv, Atal:2020igj}. However, the merger rate at high redshift is going to be large \citep{Raidal:2017mfl,Atal:2020igj} making it a key signature of PBH as dark matter. The exact dependence depends on the model for generating the PBHs \citep{Raidal:2017mfl, Vaskonen:2019jpv, Atal:2020igj}.  {For the Poisson distribution of PBHs (without clustering), the merger rate of the PBHs can be written as \citep{Raidal:2018bbj, Clesse:2020ghq}}
\begin{align}
\begin{split}\label{mergerpbh}
    \frac{R_{PBH}(z=0)}{\text{Gpc$^{-3}$ yr$^{-1}$}}= 1.6\times 10^6f_{\text{sup}}f_{\text{PBH}}^{53/37}& \eta^{-34/37}\bigg(\frac{M}{M_\odot}\bigg)^{-32/37},
    \end{split}
\end{align}
 {where $\eta\equiv m_1m_2/(m_1+m_2)^2$ is the symmetric mass ratio, $M=m_1+m_2$ is the total mass of the binaries, and  $f_{\text{sup}}$ is the suppression factor which depends on the effect from the surrounding matter distribution and also on the effects from other PBHs. One can explore both clustering and Poisson scenarios to explore the PBH population by using the evolution of the merger rate and its dependence on the population of black holes.} 

The form of the PBH merger rate as a function of redshift is fairly model-independent and can probe different populations of PBH sources. The form differs from the ABH source population, governed by the Madau-Dickinson SFR. A simple  power-law model with two free parameters, namely the power-law index $\alpha$ and the fraction of PBHs over ABHs, defined as
\begin{equation}\label{pbhsfr}
f_{PBH/ABH}\equiv \frac{\int dm_1dm_2 P_{PBH}(m_1)P_{PBH}(m_2) R_{PBH}(0, m_1,m_2)}{\int dm_1dm_2 P_{ABH}(m_1)P_{ABH}(m_2) R_{ABH}(0, m_1,m_2)}, 
\end{equation}
can cover a broad range of PBH generation scenarios \citep{Raidal:2017mfl, Raidal:2018bbj, Vaskonen:2019jpv, Atal:2020igj, DeLuca:2020qqa}. We will see that any positive value of the power-law index $\alpha$ and ratio $f_{PBH/ABH}$ can rule out the contribution of PBHs as dark matter candidates for the mass ranges accessible from the LIGO/Virgo detector network. 
The power-law functional form given in Eq. \eqref{pbhsfr} is also capable of capturing models beyond the PBH scenario such as the contribution from population-II/population-III sources below redshift $z=6$ \citep{Inayoshi:2021atf}. In future work, we will apply this technique to the population-II/population-III sources to distinguish these sources from the low-redshift ABHs and the PBH population.

With these two components, the hybrid model, including both  ABH and PBH parts,  captures both the low redshift GW sources from the stellar origin, which has a bump at a redshift $z_p$ depending on the value of time-delay, and also a monotonically increasing function which captures the redshift evolution of the PBH merger rate. 

We show in Fig. \ref{fig:mergerrate} a few examples of the hybrid merger rate for a power-law distribution of the time-delays $(t^{eff}_d)^{-1}$ with a minimum value of the time-delay parameter of $10$ Myr.  {The PBH component is shown for $f_{PBH/ABH}=1$ (by dotted lines) and $f_{PBH/ABH}=0.1$ (solid lines).} The value of the local merger rate is taken to be  $R^{GW}_0= 30$ Gpc$^{-3}$ yr$^{-1}$. The bump around redshift $z\sim 2$,  due to the peak in the star formation rate, is evident for values of $\alpha<1.28$. 
 {The power-law index of the PBH merger rate and the fraction of PBH over ABH $f_{PBH/ABH}$ determines the relative strengths of the ABH and PBH contributions if the local merger rate is fixed. So, for any large values of the parameter $\alpha>1.28$, and higher values of $f_{PBH/ABH}$, the bump due to the SFR gets obscured, whereas when the value of $\alpha$ and $f_{PBH/ABH}$ is small, the SFR bump is more prominent.} The bump moves towards a lower value of redshift if the time delay is large. The models shown here are for different values of $\alpha$, but different scenarios of PBH formation are described by the same merger rate. Larger values of $\alpha$,  with the local merger rate in agreement with the LIGO/Virgo GWTC-2 \citep{Abbott:2020niy, Abbott:2020gyp}, mimic the cases where the PBHs have large spatial clustering and the fraction of PBH as dark matter is close to unity \citep{Atal:2020igj}. Similarly  the cases with small values of $\alpha$ denote captures models with less spatial clustering and  fraction of PBH in dark matter less than $0.1$ \citep{Sasaki:2016jop, Raidal:2017mfl, Raidal:2018bbj, Young:2019gfc, Atal:2020igj,DeLuca:2020jug}. 

\begin{figure*}
    \centering
    \includegraphics[trim={0cm 0 1cm 0cm}, clip,width=1.\linewidth]{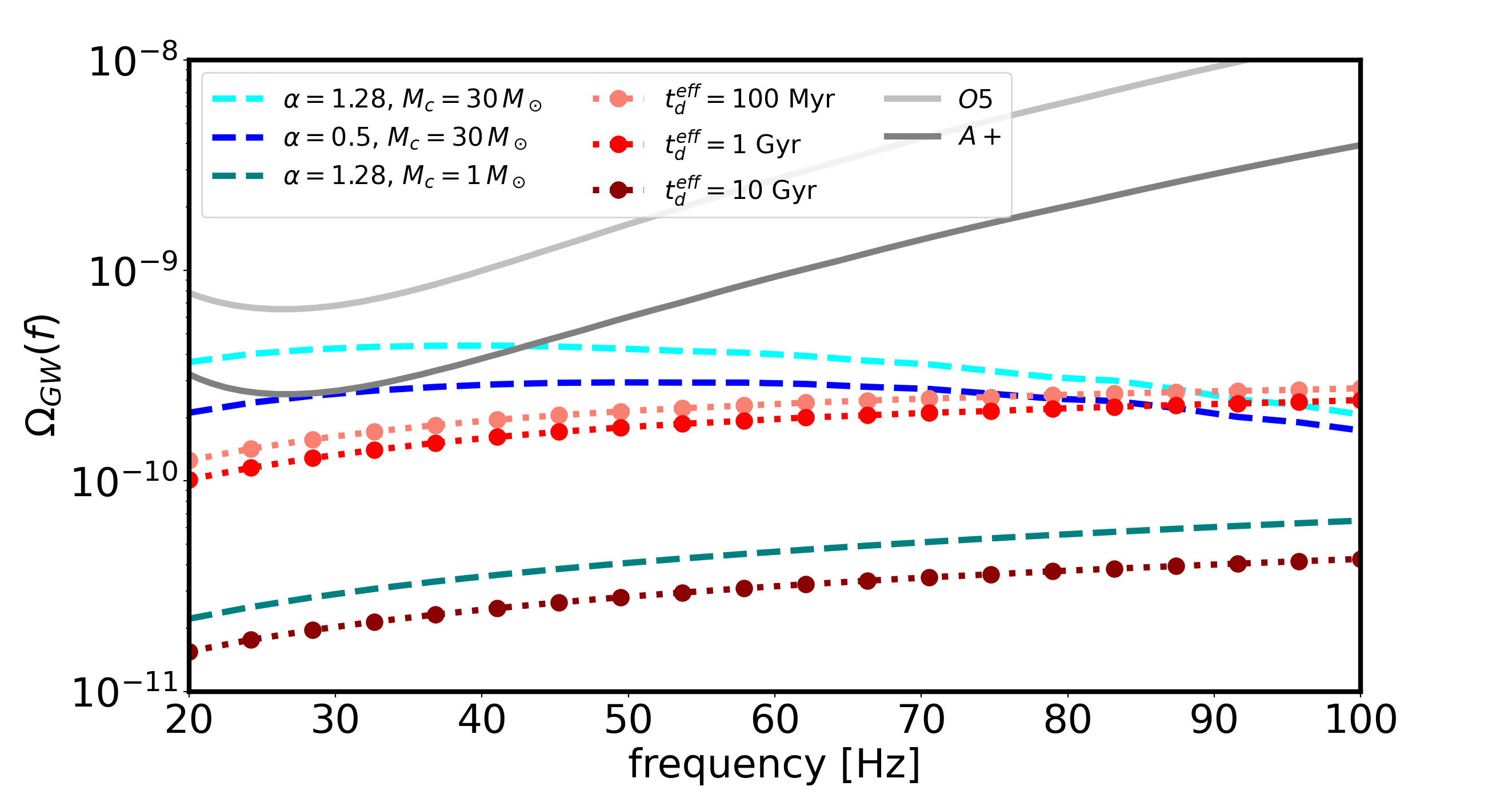}
    \caption{We show the stochastic GW background $\Omega_{GW}(f)$ as a function of frequency $f$ for   ABH sources (dotted line) for different values of the time-delay between formation  $t_d^{eff}$, and PBHs  for different values of the power-law merger index $\alpha$ and characteristic mass $M_c$. The power-law integrated noise curves for O5 and A+ are shown in light-grey and dark-grey solid lines respectively.}
    \label{fig:sgwb}
\end{figure*}

\section{Constraints using the stochastic GW background from O3}
Mergers of the GW sources at high redshift contribute to the stochastic GW background which can be written as \citep{Allen:1996vm,Phinney:2001di}
\begin{align}\label{sgwb-1}
    \begin{split}
        \Omega_{GW} (f)= \frac{f}{\rho_cc^2} \int d\theta  \int dz & \overbrace{\frac{dV}{dz}}^{cosmology}\overbrace{\frac{\mathcal{R}_{GW}(z, m_1, m_2)}{(1+z)}}^{astrophysics} \\& \times \overbrace{\bigg(\frac{1+z}{4\pi c d_L^2}\frac{dE_{GW} (\theta)}{d f_r}\bigg)}^{GW source}\bigg|_{f_r= (1+z)f}, 
    \end{split}
\end{align}
where $d_L$ is the luminosity distance,  {$\theta$ denotes the source properties such as masses of the binary black holes, their spins, inclination angle,} and $\frac{dE_{GW}}{d f_r} (\theta)$ is the energy emission per frequency bin in the source frame, written in terms of the source properties  and  chirp masses $\mathcal{M}_c$ of the gravitational wave sources as
\begin{align}\label{sgwb-1a}
    \begin{split}
        \frac{dE_{GW}(\theta)}{d f_r}= \frac{(G\pi)^{2/3}\mathcal{M}_c^{5/3}}{3} \mathcal{G}(f_r). 
    \end{split}
\end{align}
Here $\mathcal{G}(f_r)$ captures the frequency dependence during the inspiral, merger, and ringdown phases of the gravitational wave signal \citep{Ajith:2007kx}
\begin{equation}\label{fr-dep-1}
    \begin{split}
        \mathcal{G}(f_r)=
        \begin{cases}
        f_{r}^{-1/3} \, \text{for}\, f_r < f_{merg},\\  
         \frac{f_{r}^{2/3}}{f_{merg}}\, \text{for} \, f_{merg} \leq f_r < f_{ring},\\
         \frac{1}{f_{merg}f^{4/3}_{ring}}\bigg(\frac{f_{r}}{1+(\frac{f_r-f_{ring}}{f_w/2})^2}\bigg)^2\, \text{for}\,  f_{ring} \leq f_r < f_{cut},
         \end{cases}
    \end{split}
\end{equation}
where $f_{x}= c^3(a_1\eta^2 + a_2\eta +a_3)/\pi G M$ written in terms of total mass $M= m_1 + m_2$ and symmetric mass ratio $\eta= m_1m_2/M^2$. A GW binary will be emitting gravitational waves in the inspiral part up to frequency $f_{merg}$, followed by the ringdown part up to frequency $f_{ring}$, and will stop  emitting a gravitational wave signal after $f_{cut}$. $f_w$ denotes the width of the Lorentzian function.  The values of the parameters $a_1, a_2$, and $a_3$  are given in table \ref{tab:params} \citep{Ajith:2007kx}. In this analysis, we have ignored the ffect spin of the GW sources which does not make a significant difference in the GW spectrum at low frequencies \cite{Zhu:2011bd}.  

\begin{table}
    \centering
\begin{tabular}{||p{1.5cm}|p{1.5cm}|p{1.5cm}|p{1.5cm}|}
\hline
         $X_i$ &$a_1$ ($\times 10^{-1}$) &$a_2$ ($\times 10^{-2}$) &$a_3$ ($\times 10^{-2}$)\\
         \hline
         \hline
           $f_{merg}$ & $2.9740$ & $4.4810 $ & $9.5560$\\
          $f_{ring}$ & $5.9411$ & $8.9794$ & $19.111$\\
          $f_{cut}$ & $8.4845$ & $12.848$ & $27.299$\\
          $f_{w}$& $5.0801$ & $7.7515$ & $2.2369$\\
          \hline
    \end{tabular}
    \caption{We show the values of the parameters required to obtain the frequency $f_{merg}$, $f_{ring}$, $f_{cut}$, and $f_w$ denoted by the functional form $X_i= c^3(a_1\eta^2 + a_2\eta +a_3)/\pi G M$. The table is from \citep{Ajith:2007kx}.}
    \label{tab:params}
\end{table}

\begin{figure}
\centering
\subfigure[ ]{\label{fig:gauss30} 
\includegraphics[trim={0.cm 0.cm 0.cm 0.cm},clip,width=.9\linewidth]{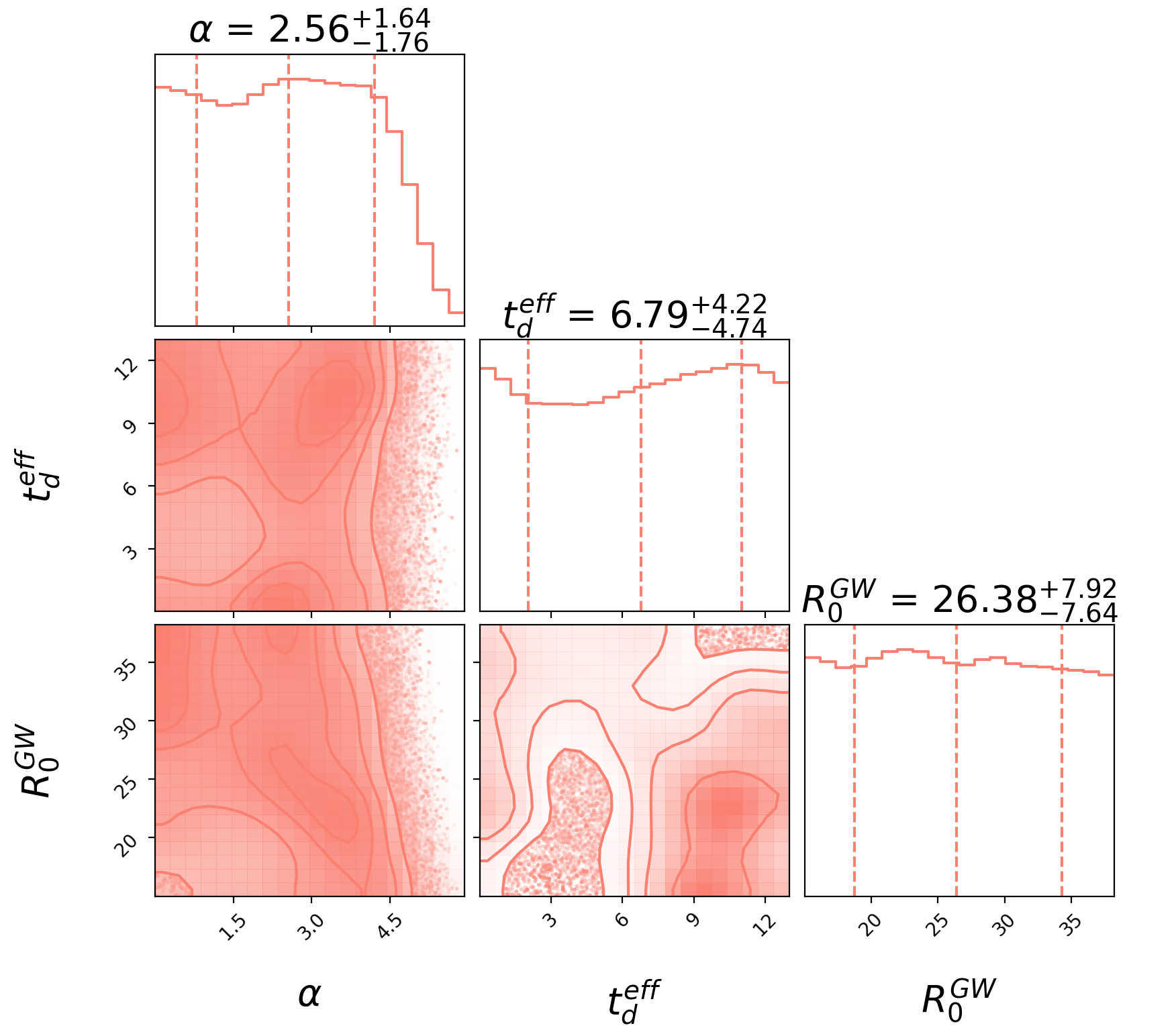}}

\subfigure[ ]{\label{fig:gauss1} 
\includegraphics[trim={0.cm 0.cm 0.cm 0.cm},clip,width=.9\linewidth]{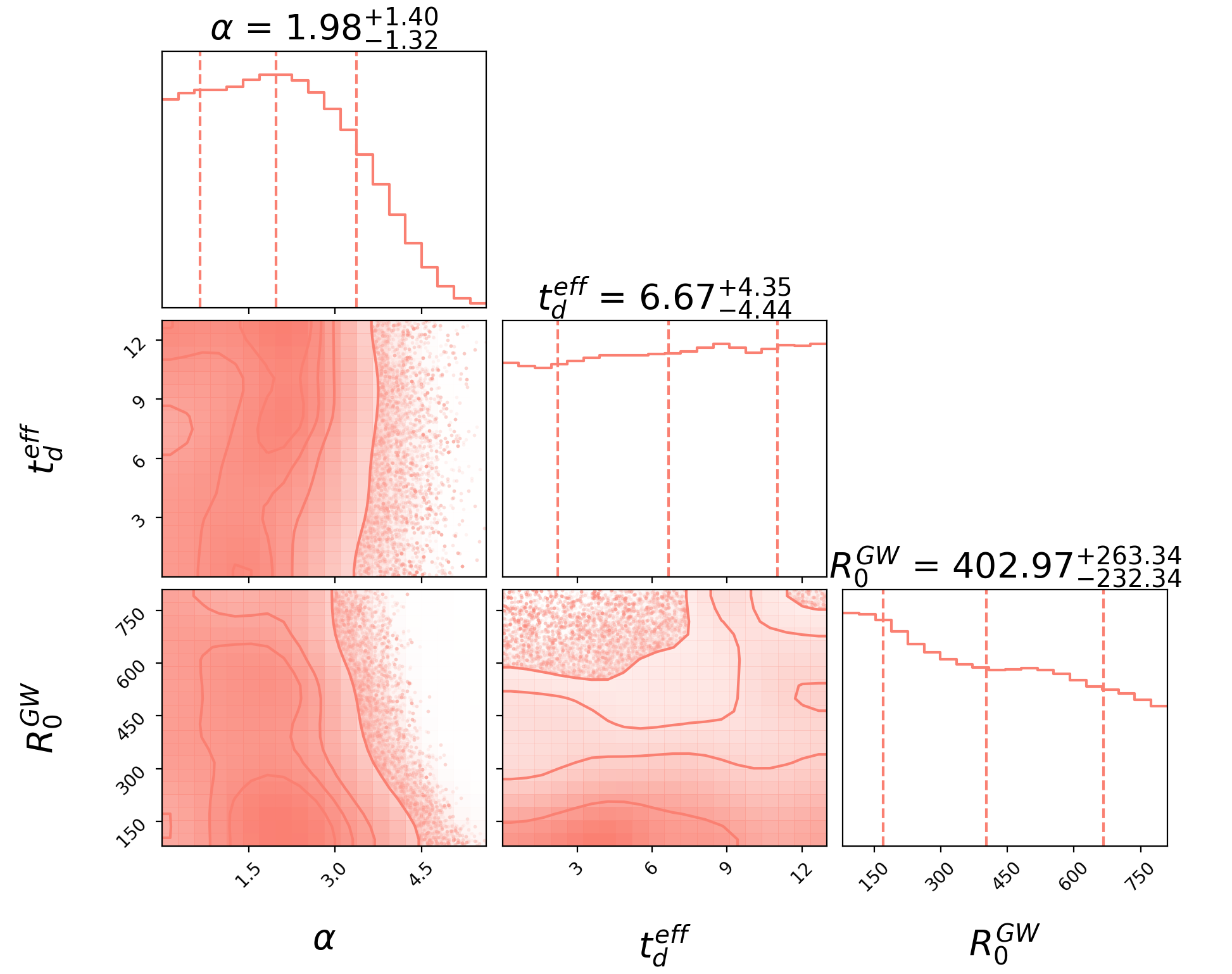}}
\subfigure[ ]{\label{fig:power-law} 
\includegraphics[trim={0.cm 0.cm 0.cm 0.cm},clip,width=.9\linewidth]{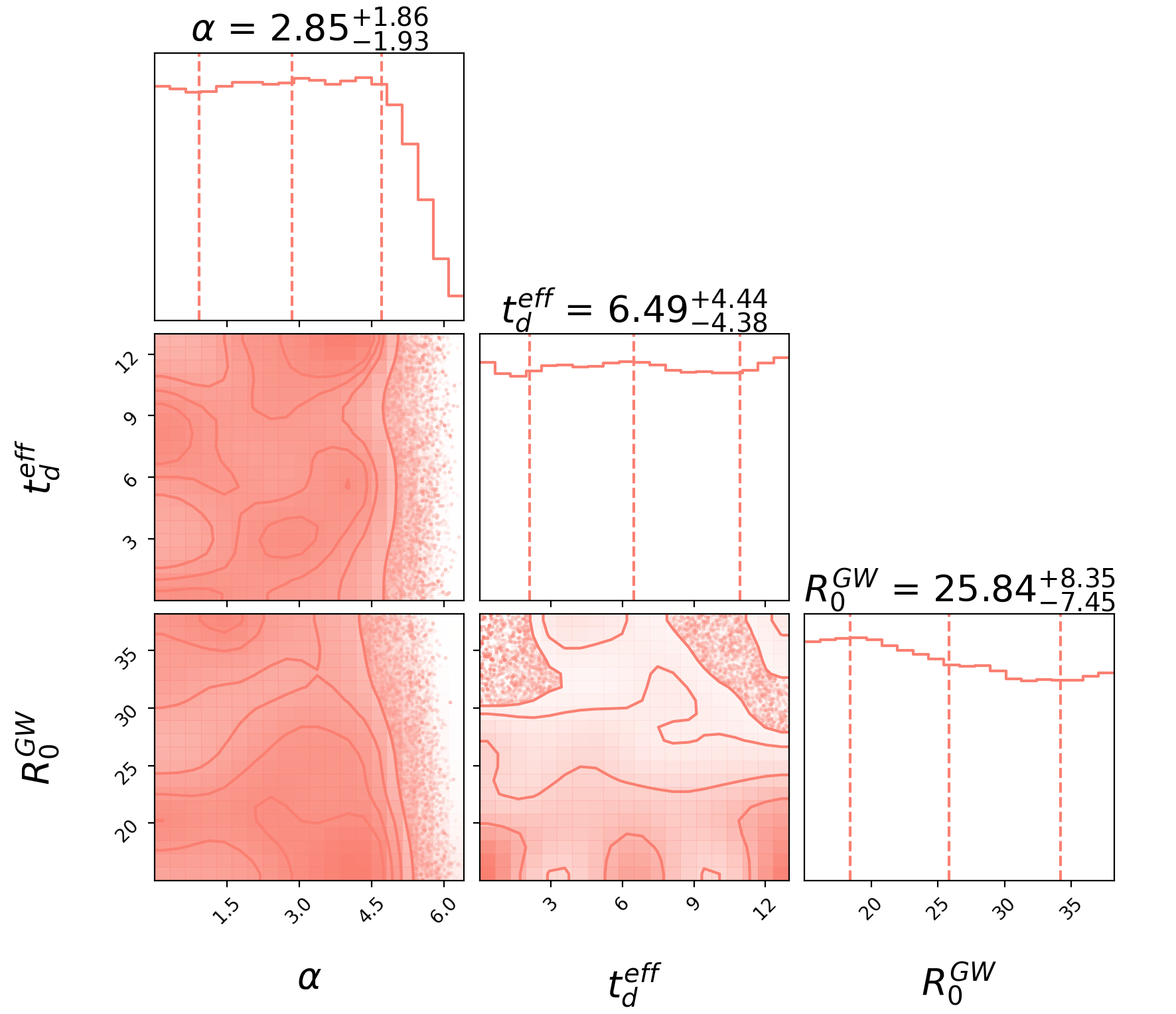}}
\captionsetup{singlelinecheck=on,justification=raggedright}
\caption{We show the joint estimation of the power-law index of the PBH merger rate $\alpha$, the time delay parameter $t_d^{eff}$, and the local merger rate $R_{0}^{GW}$ for (a) log-normal distribution with characteristic mass $M_c=30$ $M_\odot$, (b) log-normal distribution with characteristic mass $M_c=1$ $M_\odot$, and (c) power-law distribution with masses in the range $[5, \, 50]$ $M_\odot$ which is only possible for ABHs. This indicates that the current bounds presented from  O3 on these three parameters are not affected by the choice of mass-distribution.}
\label{fig:O3}
\end{figure}

We consider two different scenarios for the probability distribution of the black hole masses. For the ABHs, we consider the mass distribution of individual sources to be power-law with $m_i^{-2.3}$ for the heavier mass and flat in log-space for the lighter mass,  {following the previous stochastic analysis by the LIGO/Virgo collaboration \citep{LIGOScientific:2019vic}. One can also consider other mass distributions for estimating the stochastic background. We will show later, that the bounds obtained from the current data are not susceptible to changes in the mass distribution}. For the PBH case, we consider two different mass distributions, (i) power-law profile as for the ABH case, (ii) a log-normal mass distribution with a characteristic mass-scale $M_c$  and standard deviation $\sigma$ as
\begin{equation}
    P_{PBH}(m)= \frac{1}{\sqrt{2\pi}\sigma m}\exp\bigg({\frac{-\log^2{(m/M_c)}}{2\sigma^2}}\bigg),
\end{equation}
which is motivated by  the small-scale density fluctuations \citep{Dolgov:1992pu,Carr:2017jsz}. We consider two different values of the characteristic mass-scale $M_c$: $30 M_\odot$ and $1 M_\odot$, and the corresponding value of the parameter $\sigma$ as $0.5$.  {We show in Fig. \ref{fig:sgwb} the power spectrum of the stochastic GW background for different values of the time-delay  parameter for ABHs ($t_d^{eff}=100\, \text{Myr}, t_d^{eff}=1\, \text{Gyr}, t_d^{eff}=10\, \text{Gyr}$), and also for PBHs with characteristic mass ($M_c=30\,M_\odot$ and $M_c=1\,M_\odot$) and  power-law index ($\alpha= 1.28, 0.5$). The amplitude and shape of the stochastic GW background power spectrum varies with the changes in the model parameters. This leads to a  distinguishable signature which can be exploited to help  identify ABH and PBH sources. We also show the detector noise curve for the O5 observation run of  LIGO/Virgo \citep{TheLIGOScientific:2014jea,TheVirgo:2014hva} and for  A+ sensitivity \citep{Aasi:2013wya, aplus} in Fig. \ref{fig:sgwb}.}

We have set up a Bayesian framework (using the Bayes theorem \citep{bayes}) to estimate the parameters of the hybrid merger rate. Using the Bayes theorem, we can estimate the posterior of the power-law index $\alpha$, time-delay parameter $t^{eff}_d$, local merger-rate $R^{GW}_0$, and characteristic mass scale $M_c$ as 
\begin{align}\label{posterior}
\begin{split}
      \mathcal{P}(\vec\theta| \hat \Omega_{GW}) \propto & \mathcal{L}(\hat \Omega_{GW}|\vec\theta)\Pi(\alpha) \Pi(t^{eff}_d)\Pi(R^{GW}_0),
\end{split}
\end{align}
where $\Pi(\alpha)$, $\Pi{(t^{eff}_d)}$, and  $\Pi(R^{GW}_0)$ denote the priors on the parameters $\vec\theta\equiv \{\alpha, t^{eff}_d, R^{GW}_0\}$ and $\mathcal{L}(\hat \Omega_{GW}|\vec\theta)$ denotes the likelihood, taken as Gaussian   
\begin{align}\label{likelihoood}
\begin{split}
  \log{\mathcal{L}(\hat \Omega^{GW}_{IJ}(f)|\vec\theta )}\propto 
  \sum_{IJ, f}\frac{-\bigg(\hat \Omega^{GW}_{IJ}(f)- \Omega^{GW}(f) \bigg)^2}{2\Sigma_{IJ}(f)} ,
  \end{split}
\end{align}
where $\Omega^{GW}(f)$ is the model of the stochastic GW background signal which depends on the parameters $\alpha, t^{eff}_d,$ and $R^{GW}_0$. The measured cross-correlation signal $\hat \Omega^{GW}_{IJ}(f)= 20\pi^2f^3\text{Re}[d_I(f)^*d_J(f)]/3H_0^2T\gamma_{IJ}(f)$ between the data $d_{I,J}(f)$ from the detectors $I$ and $J$ in the Fourier domain. In this expression, $\gamma_{IJ}(f)$ denotes the overlap reduction function , $T$ denotes the observation time duration, and $H_0$ denotes the Hubble constant. The corresponding variance can be written in terms of the one-sided noise power spectrum of the detectors $P_{I,J}$ and frequency resolution $\Delta f$ as $\Sigma_{IJ}(f)= 50\pi^4f^6P_I(f)P_J(f)/9H_0^4T\Delta f\gamma^2_{IJ}(f)$.

For the analysis of the third observation run of the LIGO/Virgo \citep{Abbott:2021xxi}, we have taken  flat priors\footnote{$\mathcal{U}[a,b]$ denotes uniform distribution for the values in the range a and b.}: on the $\alpha$ parameter from $\mathcal{U}[0,10]$, and on the 
time-delay parameter $t^{eff}_d$ from $\mathcal{U}[0.01,13]$ Gyr.    We have taken a flat prior on the local merger rate parameter as $\mathcal{U}[15,38]$ Gpc$^{-3}$ yr$^{-1}$ according to GWTC-2 for GW compact objects with masses heavier than $5$ $M_\odot$ \citep{Abbott:2020niy,Abbott:2020gyp}, and for masses below $5$ $M_\odot$, we have taken a flat prior on the local merger rate as $\mathcal{U}[15,710]$ Gpc$^{-3}$ yr$^{-1}$. The contributions to the local merger rate can arise from both ABHs and PBHs, and we adopt an agnostic view with no assumptions about the separate populations. We consider that the joint contribution of the local merger rate agrees with the merger rate inferred from GWTC-2 \citep{Abbott:2020niy, Abbott:2020gyp}.  In this analysis, we consider three different models of the mass distribution, (i) log-normal distribution of the model of the  PBH masses for a fixed value of the characteristic mass $M_c=30\, M_\odot$, (ii)  log-normal distribution of the model of the  PBH masses for a fixed value of the characteristic mass  $M_c=1\, M_\odot$, and (iii) power-law mass distribution as for the ABH distribution  {with a mass range of $[5,50] M_\odot$. Though this mass cutoff at the PISN mass-scale is not motivated for any PBH mass distribution (and only possible for ABHs), we consider this to show whether the current stochastic GW observation from O3 is able to distinguish anything between the log-normal mass distribution and astrophysical mass-distribution of black holes.} For all three cases, we have considered the probability distribution of the ABH masses as power-law.  

By using the data of stochastic GW background from O3 \citep{Abbott:2021xxi} and the bounds on the local merger rate from the O3a observation \citep{Abbott:2020niy,Abbott:2020gyp}, we obtain constraints on the log-normal model of the PBH mass distribution for the value of $M_c=30\, M_\odot$ and ABH mass distribution with minimum mass $M_{min}= 5\, M_\odot$, and maximum mass $M_{max}= 50\, M_\odot$  for a power-law distribution,  shown in Fig. \ref{fig:gauss30}.  {The value of $f_{PBH/ABH}=1$ is kept fixed in this analysis. As we currently do not have any measurement of the stochastic GW signal from O3 observations, these bounds show the limits even in a  scenario when $f_{PBH/ABH}=1$, i.e. $50\%$ of the total detected black holes are of primordial origin. The bounds on the parameter $\alpha$ will get weaker if the value of $f_{PBH/ABH}<1$.} {The three upper panels of each  plot show the 1-D posteriors on the parameters $\alpha$, $t_d^{eff}$, and $R_0^{GW}$ along with the 2-D joint posteriors between the parameters in the lower panels, which is obtained using Eq. \eqref{posterior}. The plot also shows the $68\%$ and $95\%$ contours on the parameters.}
Although we do not obtain any constraint on the time-delay parameter from the non-detection of the stochastic GW background from the O3-run, there is a cut-off in the power-law index for $\alpha > 6$. This happens because, for large values of $\alpha$, the merger rates at high redshift are larger, and are constrained by the non-detection of the stochastic GW background. The constraints on the time-delay parameter are weak because the peak of the GW mergers shifts to a lower redshift in the presence of time-delay, and the relative strength at the peak is constrained by the local merger rate, as shown in Fig. \ref{fig:mergerrate}.
For the log-normal case with the PBH mass distribution having  $M_c= 1\, M_\odot$, we show the possible constraints from the O3 observation in Fig. \ref{fig:gauss1} which is similar in nature to the case $M_c=30$ $M_\odot$. The constraints on the power-law index $\alpha$ are stronger for the higher values of the local merger rate, as can be seen from Fig. \ref{fig:gauss1}. The merger rate parameters for the power-law model of the mass distribution exhibits similar constraints as for the log-normal distribution (see Fig. \ref{fig:power-law}).  {Even though for a power-law black hole mass distribution with a possible mass-cut off at the PISN mass scale (which is a possible scenario only for the ABHs) and differs from the log-normal distribution of PBH masses, the current bounds are not at all susceptible to this difference.}  From these results, we can conclude that the bounds which are obtained from the non-detection of the stochastic GW background signal from the O3 observation are nearly independent of the mass model used for PBHs.  { {The upper bound from the third observation run on the power-law index of the PBH merger rate are $\sim 2.56_{-1.76}^{+1.64}$} and the time delay parameter is $\sim 6.7_{-4.74}^{+4.22}$ Gyr at $68\%$ C.L.  {The bounds on the time-delay parameter are driven by the choice of prior (flat prior $[0.01, 13]$ Gyr) and there is no constraining power in limiting the value of time-delay from the bound on the stochastic GW background from O3 data.}
The bound on the time-delay parameter from individual events \citep{Fishbach:2021mhp} is in agreement with our result.}

\begin{figure*}
    \centering
    \includegraphics[width=1.\linewidth]{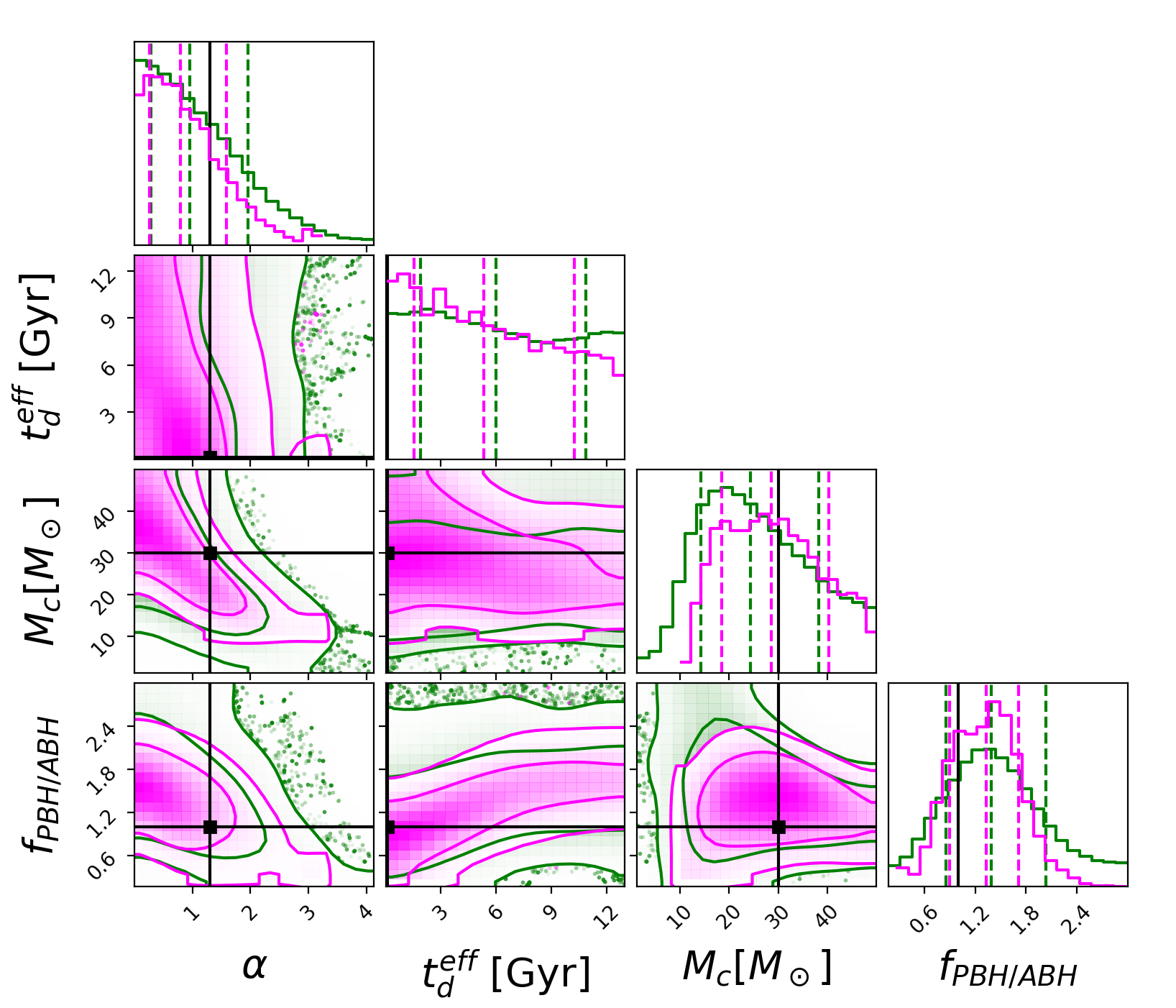}
    \caption{Forecast: we show the $68^{th}$ and $95^{th}$ contours indicating the feasibility of measuring the PBH power-law index  $\alpha=1.3$, time-delay parameter $t^{eff}_d= 100$ Myr, characteristic mass scale of PBHs $M_c=30\, M_\odot$, and the fraction  PBH/ABH $f_{PBH/ABH}=1$ from O5 sensitivity (in green) and from A+ sensitivity (in magenta). The black solid line indicates the injected value used in the simulations.}
    \label{fig:forecast}
\end{figure*}

\section{Forecast for the LIGO/Virgo and A+ sensitivity}
Although the constraints from the current LIGO/Virgo observations are weak, in the future, the stochastic GW background will become a powerful probe for distinguishing between the populations of ABHs and PBHs.  To study the feasibility of the joint estimation of the ABH  and PBH merger rates from future stochastic GW background data, we simulate the stochastic GW background signal with the hybrid model of the merger rate adopting the parameters time-delay $t^{eff}_d= 100$ Myr, $\alpha=1.3$, and  local merger rate $R^{GW}_0= 30$ Gpc$^{-3}$ yr$^{-1}$ \citep{Abbott:2020gyp}. We assume the relative fraction of sources in  ABHs and PBHs are the same ($f_{PBH/ABH}=1$). The mass distribution of the ABHs is taken to be power-law with  minimum mass $5\, M_\odot$, and maximum mass $50\, M_\odot$. The mass distribution for the PBHs is taken to be log-normal  with characteristic mass scale $M_c=30\, M_\odot$ and corresponding standard deviation $\sigma=0.5$. We consider the noise power spectrum of O5\footnote{\url{
https://dcc.ligo.org/LIGO-P1500222-v29/public}} and A+ \footnote{\url{https://dcc.ligo.org/public/0149/T1800042/004/T1800042-v4.pdf}} for this forecast \citep{TheLIGOScientific:2014jea,TheVirgo:2014hva, Aasi:2013wya, aplus}.  {The LIGO/Virgo observation at its design sensitivity for the fifth observation run with $50\%$ duty cycle is denoted by O5. The A+ sensitivity is a future upgrade of the advanced LIGO detectors almost by a factor of two \citep{aplus}. We consider the A+ noise curve integrated for two years with a duty cycle of $50\%$.}

 We consider a four-parameter model, namely the time-delay parameter $t^{eff}_d$, and the spectral index of the PBH merger rate $\alpha$,  the characteristic mass scale $M_c$, and the fraction of PBH and ABH $f_{PBH/ABH}$ as the free parameters with posterior given by
 \begin{align}\label{posterior-2}
\begin{split}
      \mathcal{P}(\vec\theta| \hat \Omega_{GW}) \propto & \mathcal{L}(\hat \Omega_{GW}|\vec\theta)\Pi(\alpha) \Pi(t^{eff}_d)\Pi(M_{c})\Pi(f_{PBH/ABH}),
\end{split}
\end{align}
 {where $\Pi(\alpha)$, $\Pi{(t^{eff}_d)}$,  $\Pi(M_c)$ and $\Pi(f_{PBH/ABH})$ denote the priors on the parameters $\vec\theta\equiv \{\alpha, t^{eff}_d, M_c, f_{PBH/ABH}\}$. For the forecast, we have taken a flat prior on the power-law index $\alpha$ from $\mathcal{U} [0,10]$, time-delay parameter $t_d^{eff}$ from   $\mathcal{U} [0.01,10]$ Gyr, characteristic mass from $\mathcal{U} [1,50]\,M_\odot$, and fraction of PBHs over ABHs $f_{PBH/ABH}$ from $\mathcal{U} [0,3]$.}
 The local merger rate is kept fixed at $R_0^{GW}=30$ Gpc$^{-3}$ yr$^{-1}$, assuming that it can be constrained from the individual detected events until the O5 and A+ runs. If there is a population of GW sources that has a merger rate that increases with redshift, then it can be isolated from the population of sources which is governed by the Madau-Dickinson law,  even if the choice of $f_{PBH/ABH}$ is kept fixed, for fast estimation of the parameters. 
 
 \begin{table}
    \centering
\begin{tabular}{||p{1.5cm}|p{2.5cm}|p{2.5cm}|}
\hline
         Parameters &O5&A+\\
         \hline
         \hline
           $\alpha$ & $(0.28, 0.95, 1.96)$ & $(0.26,  0.79, 1.59)$ \\
          $t_d^{eff}$ & $(1.88,  5.99,  10.86)$ & $(1.53, 5.34, 10.24)$\\
          $M_c$ & $(14.21, 24.37, 38.32)$ & $(18.47, 28.58, 40.27)$ \\
          $f_{PBH/ABH}$& $(0.85, 1.39, 2.04)$ & $(0.90, 1.33, 1.71)$\\
          \hline
    \end{tabular}
    \caption{We show the values of the ($16^{th},\, 50^{th},\, 84^{th}$) percentile of the forecast studies for O5 and A+ detector sensitivity on the parameters power-law index $\alpha$, time-delay parameter $t^{eff}_d$, characteristic mass $M_c$, and the fraction of PBH over ABH $f_{PBH/ABH}$. The corresponding plot of joint estimation is shown in Fig. \ref{fig:forecast}.}
    \label{tab:params-forecast}
\end{table}

 The joint estimations are shown in Fig. \ref{fig:forecast} for the O5 sensitivity in green and for the A+ sensitivity in magenta.  {The four top panels show the 1-D posteriors on the parameters $\alpha$, $t_d^{eff}$, $M_c$, and $f_{PBH/ABH}$. The joint 2-D posterior distributions between the parameters are shown in the lower panels in Fig. \ref{fig:forecast} which are obtained using Eq. \eqref{posterior-2}. We also show the $68\%$ and $95\%$ contours on the parameters in Fig. \ref{fig:forecast}.} The results show that we can measure the power-law index $\alpha=1.3$ of the PBH merger rate from future observations with O5 and A+ sensitivities. The lower values of the characteristic mass scale $M_c$ of the PBHs can be  {limited from the stochastic GW background observations for characteristic mass scale $M_c=30\, M_\odot$. However, the time-delay parameter $t^{eff}_d>100$ Myr cannot be inferred very well from the stochastic GW observations of O5 and only a partial improvement is possible from A+. For the fiducial case with a minimum time delay of $100$ Myr, from the stochastic GW background data from A+, we can expect to weakly limit the values of the time-delay parameter greater than about $9$ Gyr. Any interesting bounds from O5 on the time delay parameter are unlikely.} The parameter related to the fractional PBH and ABH $f_{PBH/ABH}$ ratio can be constrained  using the stochastic GW background data for both O5 and A+ sensitivities. This indicates that  joint estimation of the hybrid merger rates will provide an interesting avenue for distinguishing between ABHs and PBHs. We show the $16^{th},\, 50^{th},\, 84^{th}$ percentiles of the posterior distribution in Table \ref{tab:params-forecast}. Our forecast shows that  for  non-zero injected values,  the parameters $\alpha, M_c,$ and $f_{PBH/ABH}$ can be  related to the PBH fraction by these   observation run. The measurement of  non-zero values of the parameters $f_{PBH/ABH}$ and $\alpha$ from  future data on the stochastic GW background would confirm the existence of a population of black holes which are different from the population of sources that follow the  Madau-Dickinson star formation rate \citep{Madau:2014bja} in a model-independent way. Individual PBH production scenarios can be tested by using this technique.  {With a longer duration of observation time $t$, the constraints on the parameters will improve by $\sqrt{t}$.}

\section{Conclusions}
In this paper, we show how one may distinguish between different populations of GW sources using the stochastic GW background. The merger rate of the ABHs is likely to follow the star formation rate which can be modeled by the Madau Dickinson relation \citep{Madau:2014bja}, whereas, for the PBHs (or equally for  population-II/population-III sources), the merger rate is going to be different at high redshift \citep{Raidal:2017mfl, Raidal:2018bbj, Vaskonen:2019jpv, Atal:2020igj, DeLuca:2020qqa}. The merger rate of PBHs and their redshift evolution also going to depend on whether they are spatially clustered or having a Poisson distribution.    As a result,  different populations of GW sources present at early epochs can be distinguished by using the stochastic GW background to explore the high redshift Universe. {In the future a joint multi-messenger study of the stochastic GW background and different probes of star formation rate using an electromagnetic signal will further improve the capability to distinguish between the population of ABHs and PBHs.} 

We construct a hybrid model of the GW merger rates and source populations that is characterized by four parameters, namely the local merger rate of GW sources $R_0^{GW}$, the astrophysical time delay between formation and merger for ABHs $t^{eff}_d$, the power-law index of the merger rate for PBHs $\alpha$, and the characteristic mass scale $M_c$ of PBHs. We obtain  constraints on $R_0^{GW}, \,\alpha,\,t^{eff}_d$ from the stochastic GW background data of third observing run of the LIGO/Virgo collaboration O3 \citep{Abbott:2021xxi} and the bounds on the local merger rate from individual events of the O3a \citep{Abbott:2020niy,Abbott:2020gyp}   for three different mass choices (see Fig. \ref{fig:O3}). Current data can only rule out very large values of the parameter $\alpha$ and can impose no constraints on the time-delay parameter. However, in the near future from O5 and A+ sensitivities, the stochastic GW background should be a powerful probe that is capable of distinguishing between different populations of the GW sources. We show a forecast in Fig. \ref{fig:forecast}, which indicates that  bounds on the parameter $\alpha=  1.28$, and weak constraints on the time-delay parameter $t_d^{eff}=100$ Myr, characteristic mass scale of PBHs $M_c= 30$ M$_\odot$,  and the fraction of PBH/ABH $f_{PBH/ABH}=1$ should be  feasible with  O5 \citep{Martynov:2016fzi} and A+ \citep{Aasi:2013wya, aplus} sensitivities.  {Due to the correlations between the parameters, our study does not give a very large gain in constraining these parameters. But interesting conclusions on whether there exists a population of black holes with merger rate $(1+z)^\alpha$ and the possible mass-scale of these sources should be  achievable from A+, if not feasible from the O5 sensitivity.} {By using the posteriors on the PBH population parameters such as $\alpha$, $M_c$ and $f_{PBH/ABH}$, one can explore different models of PBH formation scenarios and can constrain the parameter space of those models   \citep{1967SvA....10..602Z,1971MNRAS.152...75H,1975ApJ...201....1C, 1980PhLB...97..383K, 1985MNRAS.215..575K,Carr:2005zd, Clesse:2016vqa,Ali-Haimoud:2017rtz, Raidal:2017mfl, Raidal:2018bbj,Sasaki:2018dmp,Jenkins:2020ctp,Atal:2020igj}.}    

 {In future work, we will explore the measurability of the stochastic GW signal from space-based GW detectors such as  Laser Interferometer Space Antenna (LISA) \citep{2017arXiv170200786A} and the next generation ground-based detectors such as the Einstein Telescope  \citep{Punturo:2010zz}, and the Cosmic Explorer \citep{Reitze:2019iox}. For future GW detectors, we  will be able to identify individual sources up to higher redshift ($z\sim 50$) than the detector horizon of current generation networks of detectors ($z\sim 1$). As a result, sources that are present only at a very high redshift ($z\gtrsim 50$) will contribute to the stochastic GW background signal for the next generation GW detectors. We should be able to unveil the population of high redshift GW sources.}
It should also be possible to explore the possible time dependence of the stochastic GW background \citep{10.1093/mnras/stz3226, Mukherjee:2020jxa}. This could provide an independent means  of distinguishing between different kinds of contributing sources.

\section*{Acknowledgements}
The authors are very thankful to Jishnu Suresh for reviewing the manuscript and providing very useful comments. S. M. also thanks to the group members of the LSC stochastic group for stimulating discussion on this topic. This work is part of the Delta ITP consortium, a program of the Netherlands Organisation for Scientific Research (NWO) that is funded by the Dutch Ministry of Education, Culture, and Science (OCW).  This analysis is carried out at the Horizon cluster hosted by Institut d'Astrophysique de Paris. We thank Stephane Rouberol for smoothly running the Horizon cluster. We acknowledge the use of following packages in this work: Astropy \citep{2013A&A...558A..33A,2018AJ....156..123A}, Corner \citep{corner}, emcee: The MCMC Hammer \citep{2013PASP..125..306F}, IPython \citep{PER-GRA:2007}, Matplotlib \citep{Hunter:2007},  NumPy \citep{2011CSE....13b..22V}, and SciPy \citep{scipy}. The authors would like to thank the  LIGO/Virgo scientific collaboration for providing the noise curves. LIGO is funded by the U.S. National Science Foundation. Virgo is funded by the French Centre National de Recherche Scientifique (CNRS), the Italian Istituto Nazionale della Fisica Nucleare (INFN), and the Dutch Nikhef, with contributions by Polish and Hungarian institutes. This material is based upon work supported by NSF’s LIGO Laboratory which is a major facility fully funded by the National Science Foundation.

 \section*{Data Availability}
The data underlying this article will be shared at request to the corresponding author. 

\bibliographystyle{mnras}
\bibliography{main}
\label{lastpage}
\end{document}